\pdfoutput=1
\documentclass[twoside]{article}

\usepackage{graphicx}
\usepackage{helvet}
\usepackage[T1]{fontenc} 
\usepackage[left=1.5cm, right=1.5cm, top=1.785cm, bottom=2.0cm, columnsep=20pt]{geometry}
\usepackage{multicol} 
\usepackage[labelfont=bf,up,textfont=it,up]{caption} 
\usepackage{booktabs} 
\usepackage{float} 
\usepackage{hyperref} 

\usepackage{abstract} 

\usepackage{titlesec} 
\renewcommand\thesection{\Roman{section}} 
\titleformat{\section}[block]{\large\bfseries}{\thesection.}{1em}{} 

\title{\vspace{15mm}\fontsize{20pt}{10pt}\selectfont\textbf{Plasmonic Photothermal Therapy in Third and Fourth Biological Windows}\vspace{10mm}} 
\author{\Large{E. Doruk Onal}\thanks{Corresponding author: eonal@ku.edu.tr}{ , Kaan Guven}\\[5mm] 
\large Koc University \\ 
\vspace{2mm}}

\begin{document}
\maketitle 
\begin{abstract}
{\centering
\includegraphics[scale=1.3]{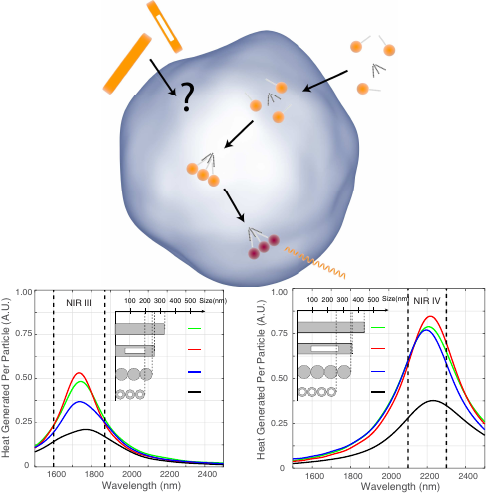}
\par
}
\vspace{5mm}
The recently reported 3rd and 4th biological transparency windows located respectively at $1.6-1.9 \mu m$ and $2.1-2.3 \mu m$ promise deeper tissue penetration and reduced collateral photodamage, yet they haven't been utilized in photothermal therapy applications. Nanoparticle based plasmonic photothermal therapy poses a nontrivial optimization problem in which the light absorption efficiency of the nanoparticle has to be maximized subject to various constraints that are imposed by application environment. Upscaling the typical absorber-dominant nanoparticle designs (rod, sphere etc.) that operate in the 1st and 2nd transparency windows is not a viable option as their size gets prohibitively large for cell intrusion and they become scatterer-dominant. The present study addresses this issue and suggests a versatile approach for designing both lithography based and self-assembling absorber dominant nanostructures for the new transparency windows, while keeping their size relatively small. The proposed nanoparticles demonstrate up to 40\% size reduction and 2-fold increase in absorption efficiency compared to the conventional nanobar design. The overall photothermal performance per nanoparticle in the 4th window is boosted up by 250\% compared to the 2nd window.
\vspace{10mm}
\end{abstract}


\begin{multicols}{2} 
\section{Introduction}

Hyperthermia therapy is based on increasing the temperature of a malignant tissue above its standard value($37^{\circ}$C) to hinder cellular processes. In this context, incorporating metallic nanoparticles (NP) that convert electromagnetic radiation into heat via plasmonic resonances has been widely investigated in the last decade and became known as plasmonic photothermal therapy (PPT). 
 
The optical response of a NP is characterized by its scattering ($\sigma_{Scat}$) and absorption cross sections ($\sigma_{Abs}$). The heating power of a metallic NP under continuous wave illumination is related to the incident light intensity and the absorption cross section of the NP: $P = I \times \sigma_{Abs}$. The efficiency of PPT is determined by these two parameters. The amount of light intensity reaching the NP is limited by the attenuation of human tissue. The absorption cross section depends on the size, shape and material of the NP. Among various materials, gold is the dominant choice for NPs especially in biological applications due to its chemical inertness, biocompatibility and ability to support localized surface plasmon resonance(LSPR). Several studies indicate that a rodlike design is the most efficient geometry for PPT applications \cite{Maestro2014, Maestro2014a, Mackey2014}. 

The light penetration problem into the human body is overcome either by using fiber optics to transmit light through the body into tumors near intrabody cavities or by utilizing light sources at certain wavelengths where human body is most transparent. These are called the biological transparency windows and located in near-infrared (NIR) region of the spectrum. So far, PPT is experimentally demonstrated in NIR-I (700-950 nm)\cite{Weissleder2001} \& NIR-II (1000-1350 nm)\cite{Smith2010} which were discovered in 2001 and 2010 respectively. Recently, advancing photodetectors and optical instruments led to the discovery of new transparency windows at longer wavelengths: NIR-III (1600-1870 nm) and NIR-IV (2100-2300 nm) in 2014 and 2016 \cite{Salas-Ramirez2014, Shi2016}. Although radiation in NIR-I \& II are successfully used in PPT applications, utilizing NIR-III \& IV assures better light penetration into deep tissue with less attenuation. To the best of our knowledge, there is no published study exploring the NPs that can operate in the NIR-III \& NIR-IV for PPT applications.

The objective of this article is to fill this gap by investigating NP designs that can
efficiently operate in these bands. As revealed in this study, upscaling the existing NP designs of NIR-I or NIR-II to the NIR-III and NIR-IV is not a feasible solution due to inefficiencies in cell intrusion and photothermal conversion.

\section{Simulation \& Modeling}

In this work we studied several NP designs from solid and contour-shaped gold nanobars to self-assembling gold nanodisk- and nanoring chains. The electromagnetic simulations are performed by a commercially available Lumerical software under linearly polarized light along the long axis of the NP. The frequency dependent complex dielectric function of gold is approximated by the Brendal-Bormann model which is shown to be in very good agreement with experimental observations in the studied wavelength range \cite{Rakic1998,Jahanshahi2014,Brendel1992}. The refractive index of the environment is set to 1.40 which corresponds to that of living cells \cite{Calin2014}.

\section{Gold Nanobars for Photothermal Therapy in NIR-III and NIR-IV}

The LSPR of a NP can be easily adjusted spectrally across the biological transparency windows by modifying its size or geometry. There is almost a linear relation between the NP length and its LSPR wavelength. However, simply scaling up the NP designs reported for NIR-I \& II is not enough to adopt them for NIR-III \& IV because of two fundamental drawbacks.

\begin{figure}[H]
    \centering
    \includegraphics*[width=82.5mm]{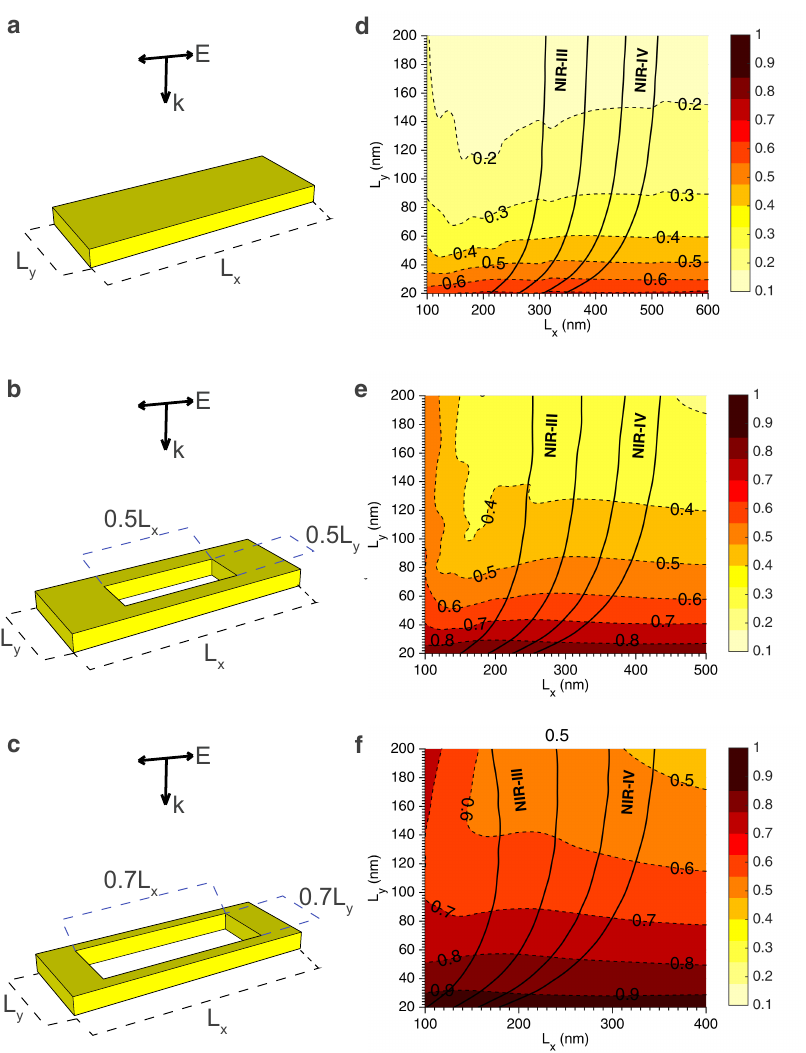}
    \caption{The schematic of (a-c) solid, 50\% contour and 70\% contour nanobars and (d-f) their respective absorption efficiency ($\phi_{Abs}$) as a function of the nanobar dimensions. Black lines indicate the regions where the nanobar is resonant in NIR-III or NIR-IV.}
    \label{fig-1}
  \end{figure}

\begin{figure*}[ht]
  \centering
  \includegraphics*[width=170mm]{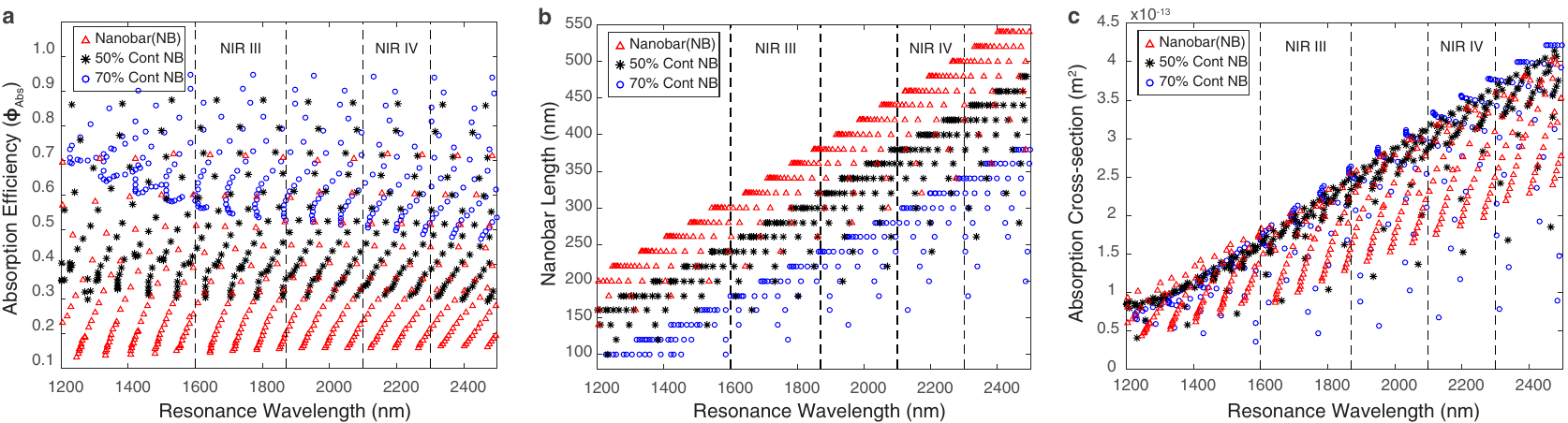}
  \caption{Benchmarking the solid nanobar (triangle) and contour nanobars (star, circle) in terms of absorption efficiency ($\phi_{Abs}$), nanobar length, and absorption cross section in NIR-III and NIR-IV. Contour nanobars provide better $\phi_{Abs}$ (a), smaller size (b), and crowd the top border of the maximum absorption cross section trend (c) across the spectra, making them suitable candidates for PPT applications.}
  \label{fig-2}
\end{figure*}

The first drawback is that scaling alters the NP's dominant response character at its LSPR:  Solid gold nanobars (or nanospheres) that are good absorbers in NIR-I become scatterer when scaled to NIR-III or IV. We show this in Fig.1.d where the absorption efficiency coefficient ($\phi_{Abs} = \sigma_{Abs} / (\sigma_{Abs} + \sigma_{Scat})$) of a solid gold nanobar (Fig.1.a) is plotted as a function of its length and width. The majority of solid nanobars for NIR-III\&IV are scatterers ($\phi_{Abs} < 0.5$). In a previous work, we proposed a contour nanobar design that enhances $\phi_{Abs}$ significantly, note how the colormapped $\phi_{Abs}$ shifts towards darker (i.e. higher) values in the entire plot region in Fig.1e\&f with increasing contour size \cite{Onal2015}. 

Even though the current research on the cell intrusion mechanism for NPs is not conclusive \cite{Alkilany2010}, decreasing the dimensions of NP would likely ease this process in addition to increasing the spatial resolution for thermal spot generation. A comparison among Figure 1.d-f also highlights that the NIR-III\&IV active regions shift towards smaller nanobar lengths with increasing contour size. Thus, the contour nanobar design aids in eliminating both of these two drawbacks simultaneously.

Increasing the absorption efficiency ($\phi_{Abs}$) is only one aspect as the absolute value of the absorption cross section ($\sigma_{Abs}$) must be taken into account in maximizing the heat generation. A recent experimental study of PPT in NIR-I employs small gold nanorods ($L = 16 - 45 nm$) that have almost $100\%$ absorption efficiency and very high cellular uptake \cite{Jia2015}. However, due to the very small $\sigma_{Abs}$, in order to generate enough heat for cell ablation, high laser intensities around $12 W/cm^2$ were required which is well above the healthy limit ($1-2 W/cm^2$). The intensity could be reduced by increasing the NP concentration but this causes further detrimental effect due to the increased cytotoxicity.

Evidently, designing NP for PPT involves many trade-offs and requires a multidimensional optimization of absorption efficiency, NP size and absorption cross section. Figure 2.a shows the contour nanobars on average achieve $100-200\%$ improvement in absorption efficiency ($\phi_{Abs} \sim 0.4 - 0.6$) depending on the contour percentage. Second, the contour nanobars provide 15-40\% reduction in size compared to solid nanobars as seen in Figure 2.b: 275-200nm vs. 350 nm in NIR III and 400-325nm vs 475nm in NIR IV, respectively. Regarding the absorption cross section, Figure 2.c shows that the contour nanobars crowd along the maximum absorption cross section trendline. This implies more freedom in design parameters while keeping the absorption cross section close to its maximum. Figure 2.c also highlights the advantages of working in longer wavelength transparency windows (i.e. NIR-III \& IV) as the maximum absorption cross section linearly increases with resonance wavelength. Therefore, heat generation per particle is significantly higher in NIR-III ($150\%$) and NIR-IV ($250\%$) in comparison to NIR-II. Overall, longer wavelength transparency windows (NIR-III \& IV) enhance the heat generation for PPT in two respects. The same amount of heat can be delivered to tumor cells at lower NP concentrations and also the energy delivery to the NPs is achieved with lower energy photons which reduces the lateral heating in healthy tissue.

As stated before, the relation between NP size and cellular uptake requires further research. However, cellular uptake for NPs that are smaller than 100 nm is well documented for PPT applications in NIR-I \& II \cite{Tsai2013, Jia2015, Mackey2014, Maestro2014, Maestro2014a}. In NIR-III \& IV even the smallest contour nanobars exceed 100 nm (Fig. 1). Fortunately, the size limitation of monolithic nanobars can be bypassed by a bottom-up approach in which disk/ring shaped NPs with diameter smaller than 100nm can be first transferred individually through the cell membrane and then assembled into a chain to construct the 'nanobar' inside the cell. We investigate this approach in the next section.

\section{Self-Assembling Nanoantenna Alternatives in NIR-III and NIR-IV: Nano Disk and Ring Chains}

The self assembly of nanodisks via DNA or protein assistance is well documented \cite{Gurunatha2016, Yin2015}. A nanodisk chain (NDC) is smaller than its monolithic nanobar counterpart but still resonates at the same wavelength \cite{Li2014a}. Replacing nanodisks by nanorings would be a simple implementation of the contour design to the self-assembling chains. There are both top-down and bottom-up methods for fabricating these nanodisk and nanoring NPs \cite{Ozel2015, Jang2014}. 

\begin{figure*}
	\centering
  \includegraphics*[width=170mm]{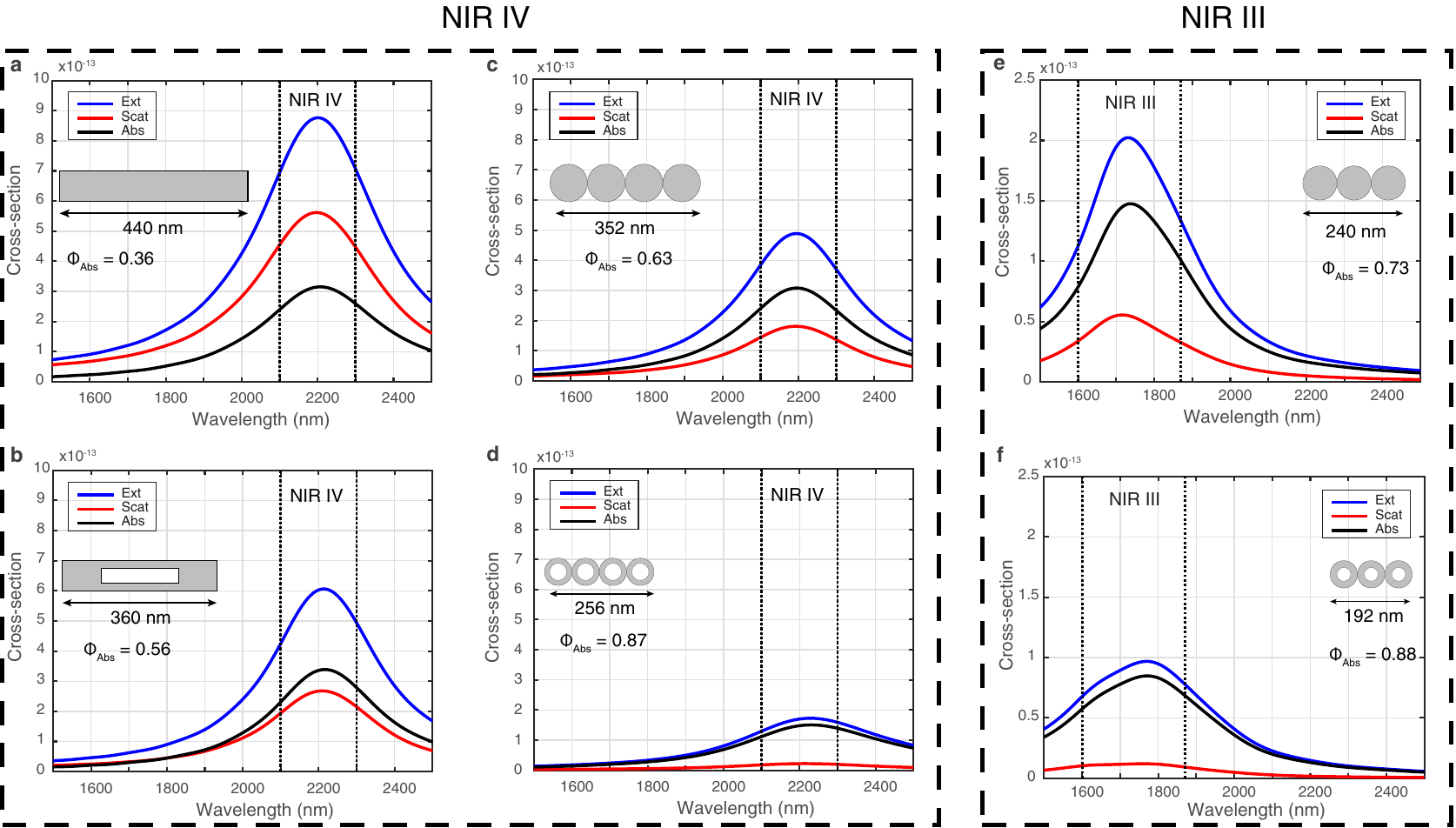}
  \caption{ A comparison of the optical cross sections (absorption, scattering and extinction) of solid nanobar, contour nanobar, NDC, NRC, designed for NIR-IV and NIR-III respectively. The contour and ring design enhances the absorption efficiency significantly.}
  \label{fig-3}
\end{figure*}

For a comparison of the absorption properties of self-assembling nanodisk/nanoring-chains and monolithic solid/contour nanobars, we picked a sample from each with the same resonance wavelength centered in the NIR-IV. The scattering and absorption cross section spectra of these samples plotted respectively in Fig. 3.a-d indicate that the NRC is the smallest in size and has the highest absorption efficiency. If we were to set the size considerations aside, the best performing NP candidate among these would be the contour nanobar with highest absorption cross section among all (7\% larger than the solid nanobar) and also providing a moderate reduction in size (18\% smaller than the solid nanobar).

The rest of the optimization for the best performing NP is about managing the trade-offs. If the future experimental evidence suggests that NPs smaller than 100 nm are the only way forward than self-assembling NP chains become the only viable option. The self-assembling structures such as NDCs come at a cost of $9\%$ reduction in absorption cross section and also equivalently in heat generated per NP. On the positive side, the absorption efficiency is increased to $63\%$ (Fig 3.c).

Replacing NDC by NRC would be only feasible when further size reduction is mandatory. This would reduce $\sigma_{Abs}$ significantly ($51\%$ compared to NDC) which effectively halves the heating power per NP. However, in photothermal imaging applications, where the maximum absorption efficiency of the NP might be of primary concern, NRCs perform better than NDCs in suppressing the scattering induced noise and interference.

Figure 3 e-f shows that choosing NIR-III as the operational window instead of NIR-IV provide further reduction in the proposed NP sizes, but also leads to smaller $\sigma_{Abs}$. It is reported that NIR-III can provide better transparency in certain tissue types compared to NIR-IV \cite{Shi2016}. Whether the increased transparency of NIR-III can compensate for the reduction of $\sigma_{Abs}$ requires further experimental data. It is also worth to point out that cytotoxic effects does not appear to be in linear relation with the NP size \cite{Alkilany2010}.

\section{Conclusion}

The contour based monolithic nanobars and self-assembling NDC/NRCs presented in this study provide both enhanced absorption efficiency and size reduction, which are the primary concerns in the optimization of NPs for PPT applications. Combined with the increased transmission and the magnitude of absorption in NIR-III\&IV, their utilization for PPT can be a viable choice. We should however point out that the relation between NP size and toxicity documented thus far in the literature introduces a nonlinear constraint to this optimization problem: A particular study reports that while small (3-5 nm) and large (50-100) NPs are not toxic, the same dose of intermediate size (18-37nm) had lethal effects on mice, linked to major organ damage \cite{Alkilany2010}.



\end{multicols}

\end{document}